\RequirePackage{fixltx2e}
\documentclass[aps,prb,reprint,floatfix]{revtex4-1}

\usepackage{amsmath,amsfonts} %
\usepackage{mathtools} %
\usepackage{wasysym} %
\usepackage{bm} %
\usepackage[bbgreekl]{mathbbol} %
\usepackage{graphicx} %
\usepackage[dvipsnames,table]{xcolor} %
\usepackage{tabularx} %
\usepackage[acronym]{glossaries} %
\usepackage{braket} %
\usepackage{fancyhdr} %
\usepackage{microtype} %
\usepackage{natmove}
\usepackage[byname]{smartref} %
\usepackage[breaklinks, %
colorlinks, linkcolor=Blue, citecolor=Blue, urlcolor=Blue]{hyperref} %
\usepackage[version=3]{mhchem}
\AtBeginDocument{\usepackage{booktabs}} 

\newcommand{\E}{\textrm{e}} %
\newcommand{\I}{\mathrm{i}\mkern1mu} %
\newcommand{\RR}{\mathbf{R}} %
\def\be{\begin{equation}}
\def\ee{\end{equation}}
\def\bea{\begin{eqnarray}}
\def\eea{\end{eqnarray}}

\def\MyTitle{Constrained variational quantum eigensolver: Quantum computer search engine in the Fock space}

\def\MyAuthora{Ilya G. Ryabinkin} %
\def\MyAuthorb{Scott N. Genin} %
\def\MyAuthord{Artur F. Izmaylov} %
\def\MySubject{Quantum computing, quantum chemistry} %

\hypersetup{ %
  pdftitle={\MyTitle{}}, %
  pdfauthor={\MyAuthora{}, \MyAuthorb{}, and \MyAuthord{}}, %
  pdfsubject={\MySubject{}}, %
  pdfkeywords={} %
}

\newacronym{FCI}{FCI}{full configurational interaction} %
\newacronym{JW}{JW}{Jordan--Wigner} %
\newacronym{BK}{BK}{Bravy--Kitaev} %
\newacronym[longplural={degrees of freedom}, %
firstplural={degrees of freedom (DOF)}, plural={DOF}]{DOF}{DOF}{degree
  of freedom} %
\newacronym[longplural={equations of motion}, %
firstplural={equations of motion (EOM)}, %
plural={EOM}]{EOM}{EOM}{equation of motion} %
\newacronym{PES}{PES}{potential energy surface} %
\newacronym{CI}{CI}{configuration interaction} %
\newacronym{VQE}{VQE}{variational quantum eigensolver} %
\newacronym{QMF}{QMF}{qubit mean field} %
\newacronym{SQP}{SQP}{sequential quadratic programming} %
\newacronym{RHF}{RHF}{restricted Hartree--Fock}

\begin{document}

\title{\MyTitle}

\author{\MyAuthora{}} %
\affiliation{Department of Physical and Environmental Sciences,
  University of Toronto Scarborough, Toronto, Ontario\, M1C\,1A4,
  Canada} %
  \affiliation{Chemical Physics Theory Group, Department of Chemistry,
  University of Toronto, Toronto, Ontario\, M5S\,3H6, Canada } %

\author{\MyAuthorb{}} %
\affiliation{OTI Lumionics Inc., 100 College St. \#351, Toronto,
  Ontario\, M5G\,1L5, Canada} %

\author{\MyAuthord{}} %
\affiliation{Department of Physical and Environmental Sciences,
  University of Toronto Scarborough, Toronto, Ontario\, M1C\,1A4,
  Canada} 
  \affiliation{Chemical Physics Theory Group, Department of Chemistry,
  University of Toronto, Toronto, Ontario\, M5S\,3H6, Canada}

  \email{artur.izmaylov@utoronto.ca}%

\date{\today}

\begin{abstract}
  Variational quantum eigensolver (VQE) is an efficient computational method promising chemical accuracy 
  in electronic structure calculations on a universal-gate quantum computer. 
  However, such a simple task as computing the electronic energy of a hydrogen
  molecular cation, \ce{H_2^+}, is not possible for a general VQE protocol 
  because the calculation will invariably collapse to a lower energy of the corresponding neutral form,
  \ce{H_2}. The origin of the problem is that VQE effectively performs an unconstrained energy optimization 
  in the \emph{Fock} space of the original electronic problem. We show how this can be avoided by 
  introducing necessary constraints directing VQE to the electronic state of interest. 
  The proposed constrained VQE can find an electronic state with a certain number of electrons, spin, 
  or any other property. The new algorithm does not require any additional \emph{quantum} resources. We demonstrate
performance of the constrained VQE by simulating various states of \ce{H_2} and 
\ce{H_2O} on Rigetti Computing Inc's 19Q-Acorn quantum processor.
\end{abstract}

\glsresetall

\maketitle



Quantum chemistry seeks the exact solution of the electronic Schr\"odinger equation,\cite{HelgakerB:2000}
\begin{equation}
  \label{eq:el_se}
  \hat H_e \ket{\Psi_i(\RR)} = E_i(\RR) \ket{\Psi_i(\RR)},
\end{equation}
where $\hat H_e$ is the electronic Hamiltonian of a molecule with a fixed nuclear configuration $\RR$, 
$E_i(\RR)$ are its eigenvalues, also known as \glspl{PES}, and $\ket{\Psi_i(\RR)}$  are the 
corresponding electronic wavefunctions. Even though this is only the 
electronic part of the total molecular quantum problem, it determines systems' properties 
crucial for designing new materials\cite{Martin:2004,Prasad:2013} and pharmaceutical compounds\cite{Roy:2015}. 
The main computational difficulty of tackling this problem 
is the exponential growth of complexity with the number of 
interacting particles ({\it i.e.} electrons). This exponential scaling 
makes it infeasible to obtain high accuracy for large systems 
({\it e.g.} materials and proteins) on a classical computer. 
Various approximations compromising the accuracy become necessary.\cite{HelgakerB:2000,Book/Parr:1989} 

On the other hand, there is a hope to overcome the exponential scaling by engaging 
a universal quantum computer.\cite{Nielsen:2010} One of the main practical difficulties remains 
maintaining large enough number of qubits in a coherent superposition state 
entangling several particles. Another issue is related to reformulating 
the electronic structure problem for the quantum computer.
The earliest proposal was the quantum phase estimation algorithm\cite{Abrams:1997/prl/2586,
  Abrams:1999/prl/5162, AspuruGuzik:2005/sci/1704}, which was quite
successful in terms of accuracy but placed strong requirements on
quantum hardware to maintain coherence for a long time. 
As an alternative with reduced coherency requirements, the \gls{VQE} has been
suggested\cite{Peruzzo:2014/ncomm/4213, OMalley:2016/prx/031007,
  Colless:2018/prx/011021}. Note though that a unitary coupled cluster type of ansatz 
  for the wavefunction used in \gls{VQE} is \emph{not} rigorously equivalent to the exact 
 solution of the electronic structure problem but rather gives numerical results of 
 chemical accuracy and is exponentially hard for a classical 
 computer. 
 
 Recent experimental work by~\citet{Kandala:2017/nature/242} demonstrated successful quantum
simulations by means of the tailored \gls{VQE} ansatz for \glspl{PES} of few selected small
molecules, \ce{H_2}, \ce{LiH}, and \ce{BeH_2}. Despite
the impressive results, there were still visible imperfections,
``kinks'', in the simulated \glspl{PES} whose origins were not clear.
The authors attributed them to the limited accuracy of simulations and
claimed that they could be removed by increasing resource requirements. 
Yet, it is still desirable to disentangle the difficulties related to
experimental realization from a possible incompleteness of
the employed formalism. One of the main goals of quantum chemistry is 
to produce smooth PESs that can be used further in 
modeling chemical dynamics. Therefore, having kinks is a significant drawback 
that cannot be left unresolved.   

Another problem that has not yet been discussed is how to apply quantum 
computing in its practical \gls{VQE} form to obtain information 
about electronic states with different numbers of electrons ({\it e.g.} cations and anions) 
or different spin ({\it e.g.} singlets, triplets, {\it etc.}). Turns out that the key
to understanding both problems, eliminating PES kinks and obtaining PESs for
different charge and spin electronic states, is the first step in the 
formulation of the electronic structure problem for quantum computing.
To encode the electronic Hamiltonian using qubits one needs to 
start not with the Hilbert space formulation \eqref{eq:el_se} but rather with 
the so-called second-quantized formulation of the Hamiltonian, which operates in the 
Fock space. The Fock space for a particular molecule 
combines the Hilbert spaces of all molecular forms with all possible number of electrons, 
This leads to an interesting problem, namely: 
for molecule \ce{A} there is only \emph{one} Hamiltonian, whose eigenvalues are electronic
energies of \ce{A}, \ce{A^{$\pm$}}, \ce{A^{2$\pm$}}, \emph{etc.} Since
the electronic energy of a cation is \emph{always} higher than that for a
neutral---otherwise a molecule would be autoionizing---it becomes an
\emph{excited} state in the full spectrum. Any variational method
aimed at \emph{minimizing} the energy will converge to the state of a
neutral \ce{A}, leaving \ce{A^{+}} inaccessible. Even worse, for any
molecule with a positive electron affinity the lowest-energy
solution is an \emph{anion} rather than the neutral form. A similar
situation is with the total electron spin, when the spin multiplicity of the
lowest energy state could be different from the one that is of interest.

We will show that in its current form, \gls{VQE} leads to kinks in PESs due to variational 
instabilities that cause switches between electronic states of different symmetries within the Fock space. 
These issues also lead to the inability of current simulation protocols to
compute \glspl{PES} of molecular cations. A particularly simple example of this problem is 
\ce{H_2^+}, which is exactly solvable problem that dates back to the early days 
of quantum mechanics\cite{Wilson:1928/prsa/635,Morse:1929/pr/932}.
We resolve all these issues by introducing the constrained modification of the \gls{VQE}, which
is indispensable for quantum chemistry applications.

\section{Results}

\subsection{Operators in the Fock and qubit spaces}
\label{sec:ferm-qubit-molec}

Formulation of the electronic structure problem for a quantum computer 
starts from the electronic Hamiltonian $\hat H_e$ in the second-quantized form
\begin{equation}
  \label{eq:qe_ham}
  \hat H_e = \sum_{ij}^{N_b} h_{ij} {\hat a}^\dagger_i {\hat a}_j + \frac{1}{2}\sum_{ijkl}^{N_b}
 g_{ijkl} {\hat a}^\dagger_i {\hat a}^\dagger_k {\hat a}_l {\hat a}_j,
\end{equation}
where ${\hat a}_i^\dagger$ and ${\hat a}_i$ are fermionic creation and
annihilation operators corresponding to a one-electron state $\phi_i$ within an 
$N_b$ one-electron basis set. 
\cite{Helgaker:2000}
The coefficients, $h_{ij}$ and $g_{ijkl}$, are one- and two-electron integrals, respectively. 
$N_b$ determines a computational cost of solving the
electronic Schr\"odinger equation because 
computational expenses grow exponentially with $N_b$ if no
approximations are made.

A Hamiltonian of a free molecule in the absence of an external
electromagnetic field forms a set of commuting operators with the
electron number operator, $\hat N$, the $z$-projection of the total
molecular spin, $\hat S_z$, and
the square of the total spin, $\hat S^2$. (The last two should not be confused 
with corresponding qubit operators.) All these operators 
have the second-quantized forms that can be found in Ref.~\citenum{Helgaker:2000}. 

Quantum computers employ two-level systems (``qubits'') as the
computational basis. Qubits can be thought of as spin-1/2 particles, 
although real quantum computers may not use 
genuine spins.\cite{Peruzzo:2014/ncomm/4213,Kandala:2017/nature/242} 
The fermionic Hamiltonian~\eqref{eq:qe_ham} is translated from fermionic to
qubit representation by one of the fermion-to-qubit
mappings, the \gls{JW} or more recent and resource-efficient
the \gls{BK} transformation~\cite{Seeley:2012/jcp/224109,
  Tranter:2015/ijqc/1431}. After the
\gls{JW} or \gls{BK} transformations all operators become
operators in the qubit space, for instance, $\hat H_e$ assumes the form
\begin{equation}
  \label{eq:spin_ham}
  \hat H = \sum_I C_I\,\hat T_I,
\end{equation}
where $C_I$ coefficients are functions of one- and two-electron integrals,  
and $\hat T_I$ operators are products of several spin operators,
\begin{equation}
  \label{eq:Ti}
  \hat T_I = \omega_0^{(I)} \cdots \omega_k^{(I)} ,\quad 0 \le k \le N_b-1.
\end{equation}
Each of $\omega_i^{(I)}$ denotes  one of the Pauli
matrices, $x_i$, $y_i$, or $z_i$ operating on the $i^{\rm th}$ fictitious spin-1/2 particle 
that represents the $i^{\rm th}$ qubit. $\hat H$ is an $N_b$-qubit operator that has a 
$2^{N_b} \times 2^{N_b}$ matrix representations.
Importantly, not only the matrix dimension but also the whole
spectrum of the qubit Hamiltonian~\eqref{eq:spin_ham} 
is identical to its fermionic counterpart~\eqref{eq:qe_ham}. Thus, finding the eigenstates of the
qubit Hamiltonian~\eqref{eq:spin_ham} is equivalent to the solving the
quantum chemistry problem.

\begin{figure}
  \centering %
  \includegraphics[width=1.0\columnwidth]{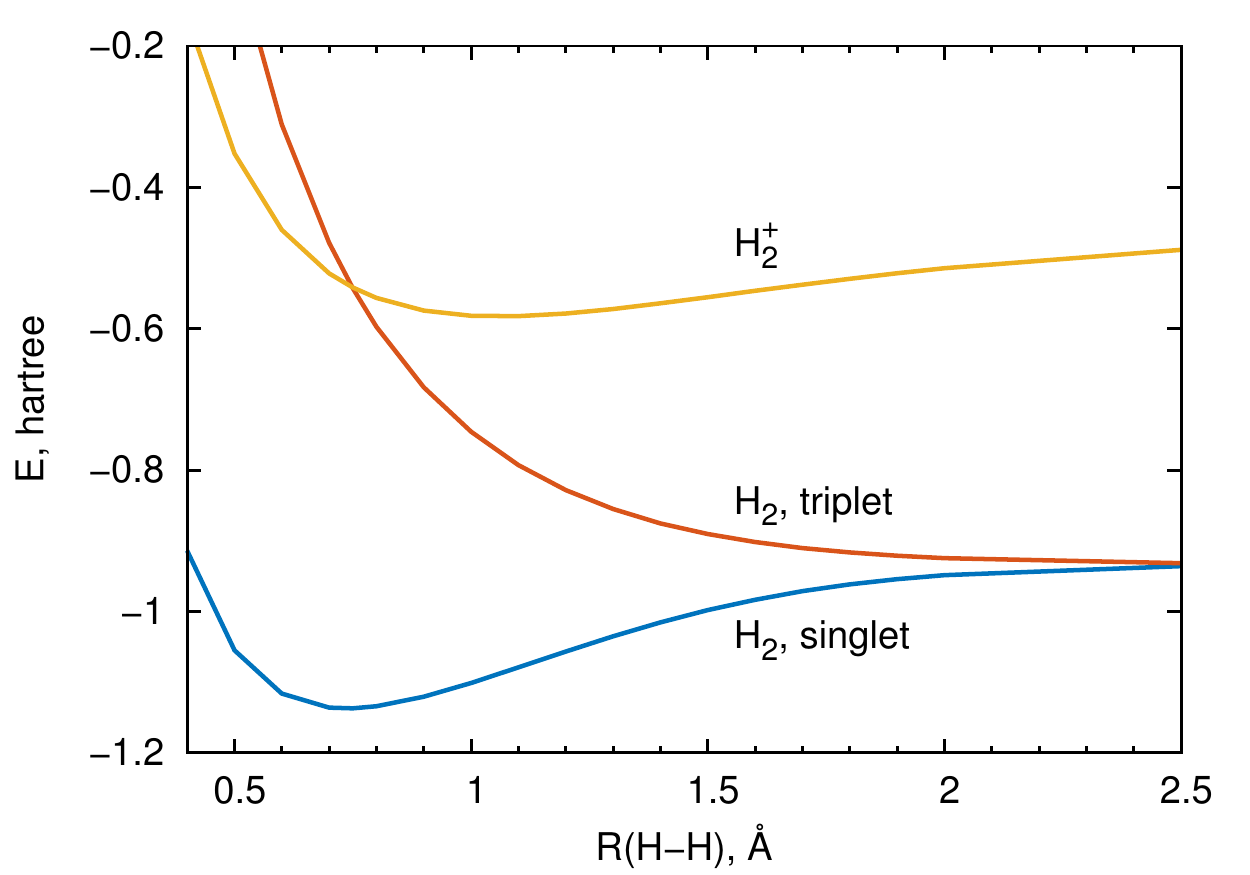}
  \caption{Two lowest PESs of the H$_2$ molecule and the ground state
  PES of the H$_2^{+}$ cation obtained using the full configuration interaction method 
  in the STO-3G basis.}
  \label{fig:h2_truth}
\end{figure}

As an illustration, we consider the \ce{H_2} molecule in the minimal STO-3G basis ($N_b=4$).    
The fermionic Hamiltonian of this system describes $2^{N_b} = 16$ electronic states.  
Presence of states of different spin and number of electrons 
does not pose a difficulty in ordinary quantum chemistry on a classical computer 
because the electronic Hamiltonian is projected onto a Hilbert subspace corresponding 
to the electronic state of interest. This projection is done via use of Slater determinants. 
They implicitly fix the number of particles $N = \braket{\hat N}$, 
$S_z = \braket{\hat S_z}$ and even $S^2 = \braket{\hat S^2}$ if appropriate 
combinations (configurations) of Slater determinants are chosen. 
Figure~\ref{fig:h2_truth} presents lowest singlet and triplet electronic PESs 
of the H$_2$ molecule and the ground state PES of its cation obtained by the 
full configuration interaction approach. 

Using the \gls{BK} transformation the electronic Hamiltonian in the STO-3G basis is mapped 
to the same number ($N_b = 4$) of qubits.
The resulting Hamiltonian\footnote{All Hamiltonians are generated using 
the OpenFermion software\cite{OpenFermion:2017}.} has 15
terms, each of them is a product of Pauli matrices multiplied by a coefficient 
inferred from one- and two-electron integrals ($h_{ij}$ and $g_{ijkl}$) at a
given interatomic distance $R$. For example, at
$R = 0.75$\AA~we have\footnote{The nuclear-nuclear
  repulsion energy $1/R = 0.705333$ a.u. is included in this Hamiltonian.}
\begin{equation}
  \label{eq:H2_sto3g_0.75A}
  \begin{split}
    & \hat H_\text{BK}(R)\Big|_{R =0.75\AA} =
    - 0.109731  \\
    & + 0.169885\,z_0\phantom{z_2z_1z_0} + 0.168212\,z_1\phantom{z_2z_1z_0} \\
    & + 0.169885\,z_1z_0\phantom{z_2z_1} + 0.0454429\,x_2z_1x_0 \\
    & - 0.218863\,z_2\phantom{z_2z_1z_0} + 0.0454429\,y_2z_1y_0 \\
    & + 0.120051\,z_2z_0\phantom{z_1z_0} + 0.165494\,z_2z_1z_0\phantom{z_2} \\
    & + 0.173954\,z_3z_1\phantom{z_2z_1} + 0.0454429\,z_3x_2z_1x_0 \\
    & + 0.120051\,z_3z_2z_0\phantom{z_2} + 0.0454429\,z_3y_2z_1y_0  \\
    & - 0.218863\,z_3z_2z_1\phantom{z_2} + 0.165494\,z_3z_2z_1z_0.
  \end{split}
\end{equation}
Diagonalization of this Hamiltonian in the $2^{N_b} = 16$-dimensional qubit space provides the same
eigenvalues, but the information about the number of electrons $N$, or $S^2$ for corresponding 
eigenstates is now hidden. 

\begin{figure}
  \centering %
  \includegraphics[width=1.0\columnwidth]{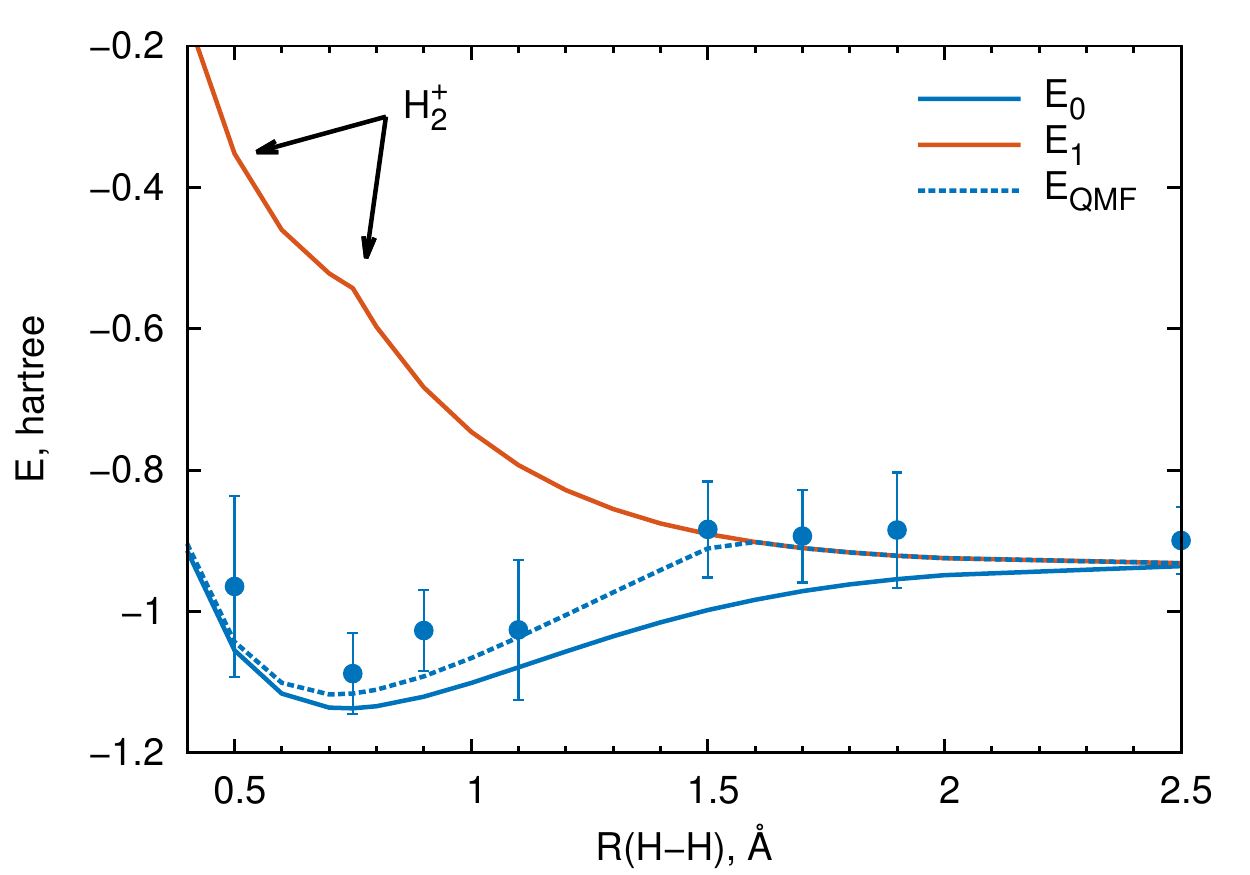}
  \caption{Two lowest eigenstates of the
    Hamiltonian $\hat H_{\rm BK}(R)$ (solid lines), and the PES 
    corresponding to the minimum of the \protect\gls{QMF} functional,
    Eq.~\eqref{eq:mf_func_def} (dashed line).
    Points with error bars correspond to \protect\gls{QMF} calculations performed on the Rigetti 
    quantum computer. Error bars show the standard deviation over measured values.}
  \label{fig:h2_exact_vs_qmf_unconstr}
\end{figure}

Let us consider the two lowest \emph{exact} PESs 
of the \ce{H_2} molecule that were calculated by full diagonalization of the qubit
Hamiltonian $\hat H_{\rm BK}(R)$ for different $R$ (Fig.~\ref{fig:h2_exact_vs_qmf_unconstr})
We track the physical nature of the solutions using properties corresponding 
to commuting observables: first, we constructed the
\gls{BK}-transformed operators $\hat N$ and $\hat S^2$
\bea \label{eq:hatN_BK}
  \hat N_\text{BK} &=& 2 - (z_0 - z_1z_0 - z_2 - z_3z_2z_1)/2, \\
  \notag
  \hat S^2_\text{BK} &=&  (6 - 3z_1 + x_2x_0 - x_2z_1x_0 + y_2y_0 \\ \notag
  &+& z_2z_0 - z_2z_1z_0 - 3z_3z_1 - y_2z_1y_0 \\ \notag
  &+& z_3x_2x_0 - z_3x_2z_1x_0 + z_3y_2y_0 \\ 
  &-& z_3y_2z_1y_0 + z_3z_2z_0 - z_3z_2z_1z_0)/8
\eea
and then evaluated the mean values of these operators on the
calculated exact states. First of all, for $R \le 0.7$\AA~the first excited state
corresponds to the state of \ce{H_2^{+}}, $N = 1$, while for larger
$R$ it is a triplet state of the neutral molecule ($N = 2$, $S^2 = 2$).
Thus, in the STO-3G basis set the triplet state changes its
position in the spectrum as $R$ varies.
The cationic ground state is among the excited states and intersects with the triplet H$_2$ state
forming a kink due to the energy ordering of the electronic states. Therefore, 
one of the reasons for appearing kinks in quantum calculations can be intersections of states that 
originally belonged to different Hilbert spaces of the fermionic problem and brought within the same 
qubit space by using the Fock-space second-quantized Hamiltonian \eqref{eq:qe_ham}.


\subsection{Variational quantum eigensolver}
\label{sec:appr-meth-solv}

VQE carries out the optimization of the electronic ground state energy in a two-step
procedure. First, the expectation value of the Hamiltonian
\bea
  \label{eq:barE}
  E(\boldsymbol \Omega, \boldsymbol\tau) = \braket{\Psi(\boldsymbol \Omega,\boldsymbol\tau)|\hat
    H|\Psi(\boldsymbol \Omega,\boldsymbol\tau)}
\eea
was calculated from measurements of individual Pauli terms, $\hat T_I$
[Eq.~\eqref{eq:spin_ham}], to obtain corresponding averages  
$\braket{\Psi(\boldsymbol \Omega,\boldsymbol\tau)|\hat T_I|\Psi(\boldsymbol \Omega,\boldsymbol \tau)}$
and to contract them with $C_I$ at fixed values of wavefunction parameters
$\boldsymbol \Omega,\boldsymbol \tau$. 
Second, the minimization of $E(\boldsymbol \Omega,\boldsymbol \tau)$ is done 
on a classical computer  
\bea
E = \min_{\boldsymbol \Omega,\boldsymbol \tau} E(\boldsymbol \Omega,\boldsymbol \tau). 
\eea
A typical parametrization of the wavefunction as a $N_b$-qubit trial state is
\bea
  \label{eq:ibm_vqe_ansatz}
    \ket{\Psi(\boldsymbol \Omega,\boldsymbol \tau)} &= & U_\text{ENT}(\boldsymbol \tau) U_\text{MF}(\boldsymbol \Omega) \ket{00 \ldots 0},
\eea
where $\ket{00 \ldots 0}$ is an initialized $N_b$-qubit state, 
$U_\text{MF}(\boldsymbol \Omega)$ is a mean-field rotation of individual qubits and 
$U_\text{ENT}(\boldsymbol \tau)$ is an ``entangler'' that is responsible for post mean-field 
treatment of electron-electron correlation effects. 
Entanglers are parametrized as an exponent of multi-qubit 
anti-hermitian operators that have parametric dependence on components of $\boldsymbol \tau$. 

For individual qubit rotations in $U_\text{MF}(\boldsymbol \Omega)$ of Eq.~\eqref{eq:ibm_vqe_ansatz}, 
only two out of the three Euler angles change the total energy,
and one angle defines a global phase change, which does not affect the
energy. A convenient basis for the Hilbert space of individual qubits representing these relations 
is a basis of spin coherent states~\cite{Perelomov:1972, Radcliffe:1971/jpa/313,
  Arecchi:1972/pra/2211, Lieb:1973/cmp/327}, $\{\ket{\Omega}\}$, where
$\Omega = (\phi, \theta)$ encodes a position of a qubit orientation on the 
Bloch sphere (see Methods for more formal definitions).
The direct product of individual qubit coherent states forms the $N_b$-qubit mean-field solution 
$\ket{\boldsymbol \Omega} = U_\text{MF}(\boldsymbol \Omega)\ket{00 \ldots 0}$. 
The optimal values of $\boldsymbol \Omega$ can be obtained using the variational principle,\cite{Lieb:1973/cmp/327}  
if $E_0$ is the ground state energy for the Hamiltonian in Eq.~\eqref{eq:spin_ham} 
then $E_0 \le \braket{\boldsymbol \Omega | \hat H | \boldsymbol \Omega }=E_\text{QMF}(\boldsymbol \Omega)$.
Therefore, $\braket{\boldsymbol \Omega | \hat H | \boldsymbol \Omega }$ defines the
\emph{qubit mean-field energy functional},
\begin{equation}
  \label{eq:mf_func_def}
  E_\text{QMF}(\boldsymbol \Omega) = \sum_I C_I F_I(\mathbf{n}_1^{(I)}, \ldots
  \mathbf{n}_{N_b}^{(I)}),
\end{equation}
where each $F_I$ is obtained from $\hat T_I$ by substitution
$\omega_i^{(I)} \to \mathbf{n}_i^{(I)}$ 
and operator products of $\omega_i^{(I)}$ are converted to ordinary
numerical products. $\mathbf{n}_i^{(I)}$ is a shorthand notation
for the $\omega_i^{(I)}$ component of the unit vector on a Bloch sphere:
$\mathbf{n} = (\cos\phi\sin\theta,\, \sin\phi\sin\theta,\,
\cos\theta)$.

The ground-state mean-field solution for H$_2$, $E_\text{QMF}$ 
(Fig.~\ref{fig:h2_exact_vs_qmf_unconstr}), which behaves 
like the \gls{RHF} curve for small $R$, has a second type of kinks near
$R \approx 1.6$\AA. This kink is due to switching of
a mean-field minimum from a singlet ($S^2 = 0$) to a
triplet ($S^2 = 2$) solution. 

To expose the second type of kinks in quantum computing for PESs in their most vivid form 
we avoided using entanglers. Generally, entanglers are supposed to bring the mean-field 
solution closer to the exact one. The later is smooth in this case (Fig.~\ref{fig:h2_exact_vs_qmf_unconstr}),
and therefore, if the entanglers fully accomplished the task there would be no kink.
However, in practical calculations there is no general prescription how to choose the entangler 
that rigorously guarantees convergence to the exact answer. Thus, any approximate
entangler can make kinks originating in mean-filed solutions less pronounced but still existent.   

\subsection{Constrained Variational Quantum Eigensolver}

To modify the variational procedure to
include information about $N$ and $S^2$ of the target state we use the
\emph{constrained optimization}. A constrained minimization 
is readily applicable to \gls{VQE} by adding a penalty functional
\bea
  \label{eq:EcG}
 \mathcal{E}(\boldsymbol \Omega, \boldsymbol \tau, \boldsymbol \mu)
  & = & \braket{\Psi(\boldsymbol \Omega, \boldsymbol \tau)|\hat
    H|\Psi(\boldsymbol \Omega, \boldsymbol \tau)}  \\
    &&+ \sum_i\mu_i
    \left[\braket{\Psi(\boldsymbol \Omega, \boldsymbol \tau)|\hat C_i 
    |\Psi(\boldsymbol \Omega, \boldsymbol \tau)} - C_i\right]^2, \notag
\eea
where $\hat C_i$ are constraining operators ({\it e.g.} $\hat N$, $\hat S^2$, {\it etc.}),
$C_i$ are their desired mean values, and $\mu_i$ are big but fixed
numbers.\cite{Nocedal:2006} Comparing operators in Eqs.~\eqref{eq:H2_sto3g_0.75A} 
and~\eqref{eq:hatN_BK} one can see that they share the same Pauli
terms, which means that the value
$\braket{\Psi(\boldsymbol \Omega, \boldsymbol \tau)|\hat C_i|\Psi(\boldsymbol \Omega, \boldsymbol \tau)}$
can be formed reusing values of $\braket{\Psi(\boldsymbol \Omega, \boldsymbol \tau)|\hat T_I|\Psi(\boldsymbol \Omega, \boldsymbol \tau)}$ 
at zero additional cost.

\begin{figure}
  \centering %
  \includegraphics[width=1.0\columnwidth]{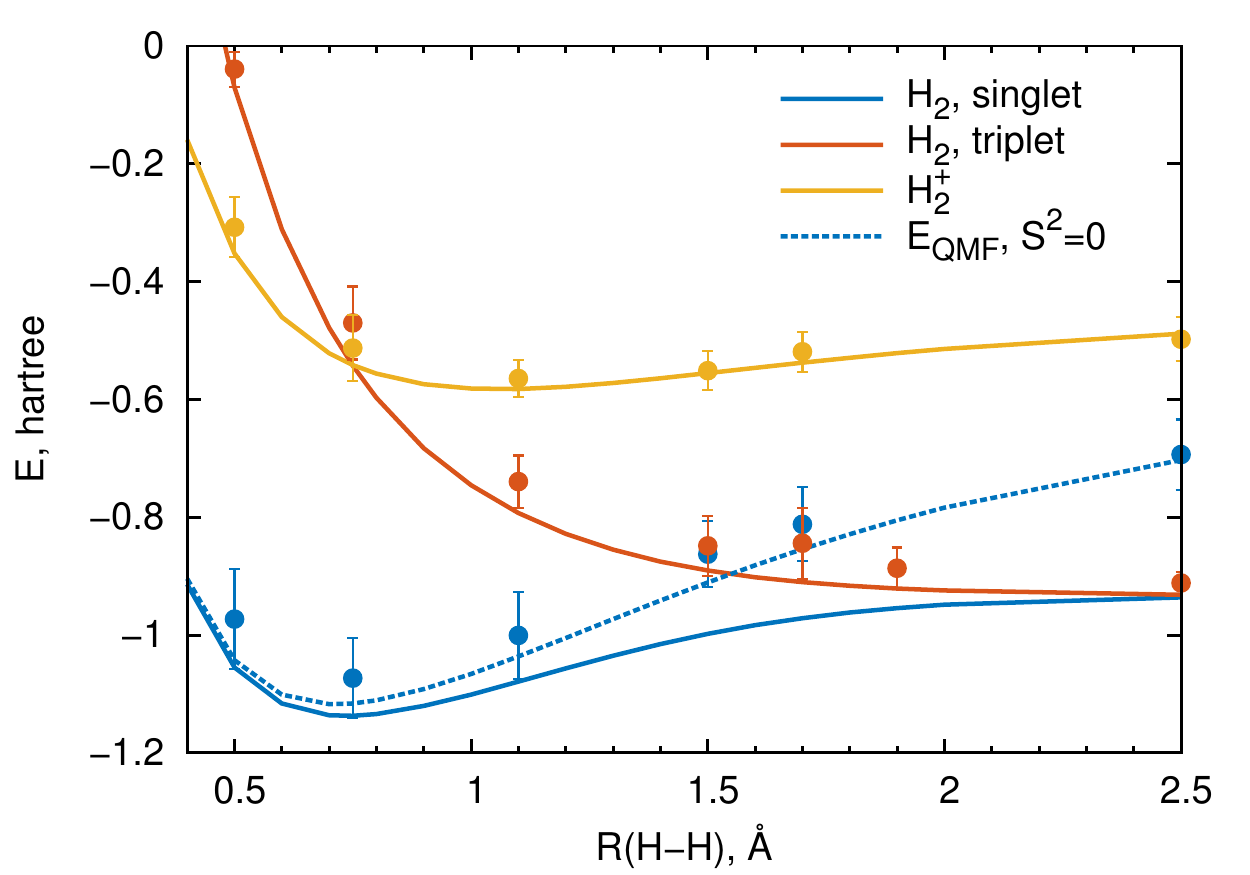}
  \caption{Three constrained mean-field PESs for lowest singlet ($S^2=0$) and triplet ($S^2=2$) electronic 
  states of H$_2$ and the ground electronic state of H$_2^{+}$ ($N=1$). The exact PESs for all but the singlet 
  H$_2$ state coincide with the constraint mean-field solutions. 
  Points with error bars correspond to constrained \protect\gls{QMF} calculations 
  performed on the Rigetti quantum computer.
    Error bars show the standard deviation over measured values.}
  \label{fig:h2_constr}
\end{figure}

To obtain lowest energy mean-field solutions of neutral \ce{H_2} with well-defined electron spin 
we minimize the following functional
\begin{align}
  \label{eq:cS}
  \mathcal{E}_S(\boldsymbol \Omega, \mu)
  & = E_\text{QMF}(\boldsymbol \Omega) + \mu \left[\braket{\boldsymbol \Omega |\hat S_\text{BK}^{2}|\boldsymbol \Omega} - S^2\right]^2,
\end{align}
where $S^2$ is either 0 (singlet) or 2 (triplet). 
%
The PESs of such constrained minimizations do not exhibit any kinks and retain their 
target spin values at all $R$ (Fig.~\ref{fig:h2_constr}).
The constrained mean-field singlet PES demonstrates the same 
asymptotic behavior as the \gls{RHF} curve by going to the incorrect
dissociation limit that is exactly in between purely covalent H$\cdot$ + H$\cdot$
and ionic H$^{+}$ + H$^{-}$ solutions. To correct for this behavior requires 
an addition of an entangler. On the other hand, the triplet mean-field 
counterpart reproduces the exact triplet PES because there is no 
electron-electron correlation for the triplet state in this minimal basis setup. 

Similarly, the constrained minimization of the functional 
\begin{align}
  \label{eq:cN}
  \mathcal{E}_N(\boldsymbol \Omega, \mu)
  & = E_\text{QMF}(\boldsymbol \Omega) + \mu \left[\braket{\boldsymbol \Omega |\hat N_\text{BK}|\boldsymbol \Omega} - 1\right]^2
\end{align}
that imposes the $N = 1$ constraint has been employed to extract 
the lowest PES of \ce{H_2^{+}} (Fig.~\ref{fig:h2_constr}).
The resulting curve is smooth and coincides with the
 exact H$_2^{+}$ PES due to the absence of electron-electron correlation. 

\begin{figure}
  \centering %
  \includegraphics[width=1.0\columnwidth]{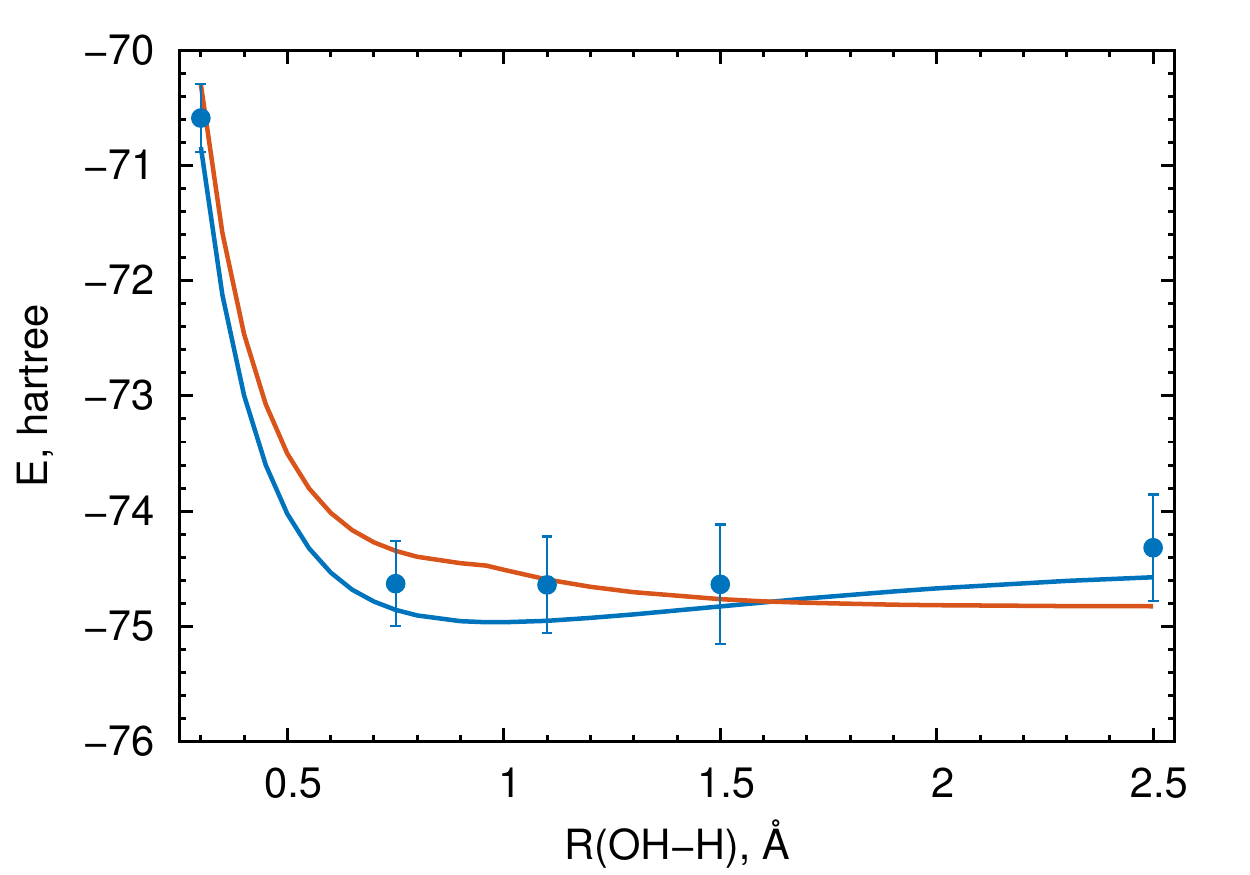}
  \caption{Restricted Hartree--Fock PESs for the lowest singlet (blue line) and triplet (red line)
  electronic state of H$_2$O obtained on a classical computer. Points with error bars correspond to 
  the constrained \protect\gls{QMF} calculations for the lowest singlet state 
  performed on the Rigetti quantum computer. Error bars show the standard deviation over measured values.}
  \label{fig:h2o}
\end{figure}

The constrained methodology can be easily extended to larger systems 
where constraints become especially useful due to increasing density of electronic states.
As an example, we consider the ground singlet state of the water molecule in the STO-3G basis. 
Figure~\ref{fig:h2o} shows its PES obtained by changing a single OH-bond length from 
the symmetric equilibrium configuration with $R$(OH)$=0.9605$\AA~and $\angle$HOH$=104.95^{o}$.
Here, we used both spin ($S=0$) and number of electrons ($N_e=10$) constraints to obtain 
the mean-field solution. The spin constraint is invaluable for avoiding 
convergence to the closely spaced triplet solution (Fig.~\ref{fig:h2o}).  

Another advantage of using the constraints in quantum computing is their 
capacity to reduce the noise in the measurement process.\cite{Rubin:2018tf} 
Generally, there are two sources of uncertainty when the Hamiltonian components 
are measured in the quantum computer: 1) the wavefunction encoded in qubits 
is not an eigenfunction of a particular Pauli word ($T_I$) and hence 
there is an intrinsic quantum uncertainty for the $T_I$ measurement, 
2) uncontrolled interactions of qubits with their environment that introduce all sorts of noise 
unrelated to the ideal, quantum uncertainty. A typical measurement on a quantum computer 
produces an eigenstate of a measured Pauli word, $T_I$. 
We found that re-weighting results 
of the measurement based on overlaps of a collapsed wavefunction with 
eigenfunctions of the property operators ({\it e.g.} $\hat N$ and $\hat S^2$) 
with target eigenvalues (fixed charge and spin) strongly reduces the noise coming from the 
second source, and does not alter the statistics originating from the truly quantum uncertainty.
Our constraint based post-processing scheme is detailed in the Methods section. 


\subsection{Discussion}
\label{sec:conclusions}

We have proposed a simple constrained VQE approach that is \emph{indispensable} 
if one seeks a solution of a quantum chemistry problem for an electronic state with 
a well-defined electronic spin, charge, or any other property of interest. 
The corresponding procedure requires minimal modification
of the current \gls{VQE} protocol and incurs virtually no additional
\emph{quantum} costs. Using the constrained VQE not only allows one
to target specific states but also removes kinks in PESs arising due to numerical 
instabilities associated with the root switching.  

In the current study only the electron number and the total spin operators have been used 
for imposing constraints. The $z$-projection of the total electron spin ($\hat S_z$) 
has not been constrained, although it is another symmetry that one can use. 
We found that restriction of $S_z$ was not giving any improvements 
in the considered cases. 


Moreover, the operators of conserved quantities
can be used to post-process the results of measurements done on the quantum computer
to reduce the noise due to qubit interactions with the environment. This post-processing does 
not affect the statistics of the true quantum distribution arising from the ideal projective 
measurement.  

\section{Methods}


\subsection{Spin coherent states}
\label{sec:spin-coherent-states}

A spin coherent state, also known as a ``Bloch state'', for a single
particle with spin $J$ ($J \ge 0$ is integer or half-integer) is
defined by the action of an appropriately scaled exponent of the
lowering operator $\hat S_{-}$ on the normalized eigenfunction of
$\hat S_z$ operator, $\hat S_z \ket{JM} = M \ket{JM}$, with maximal
projection $M = J$\cite{Lieb:1973/cmp/327}:
\begin{align}
  \label{eq:spin_coh_state}
  \ket{\Omega} = & \cos^{2J}\left(\frac{\theta}{2}\right) \exp\left[\tan
                   \left(\frac{\theta}{2}\right)\, \E^{\I\phi}\, \hat S_{-}\right]
                   \ket{JJ} \nonumber \\
  = & \sum_{M = -J}^J \binom{2J}{M+J}^{1/2} \nonumber \\
                 &\times  \cos^{J+M}{\left(\frac{\theta}{2}\right)}
                   \sin^{J-M}\left(\frac{\theta}{2}\right)
                   \E^{\I(J-M)\phi} \ket{JM},
\end{align}
where the $\ket{JM}$ states are normalized as
\begin{equation}
  \label{eq:smin_act}
  \ket{JM} = \binom{2J}{M+J}^{1/2} [(J-M)!]^{-1} S_{-}^{J-M} \ket{JJ}.
\end{equation}
$\{\ket{\Omega}\}$ constitute an overcomplete non-orthogonal set of
states on a unit Bloch sphere parametrized by spherical angles
$\Omega = (\phi, \theta)$, $0 \le \phi < 2\pi$, $0 \le \theta \le \pi$.

\subsection{Classical minimization}

Constrained [Eqs.~\eqref{eq:cS} and \eqref{eq:cN}] and unconstrained [Eq.~\eqref{eq:mf_func_def}] 
minimizations on a classical computer were done 
using the \gls{SQP} algorithm as implemented by the \texttt{fmincon} routine of the
\textsc{matlab}\cite{MATLAB:2015} software.
 All mean-field solutions were obtained by minimizing the corresponding energy functions 
 with respect to all $4 \times 2 = 8$ Bloch angles. 

\subsection{Quantum computer simulation details}
\label{sec:quant-comp-simul}

We performed the simulations on the Rigetti 19Q-Acorn quantum processor unit 
(QPU)\cite{Otterbach:2017uh} 
using pyQuil and Forest API.\cite{Smith:2016tp} 
Our wavefunction ansatz was obtained performing RZ and RX gate operations on individual qubits. 
This corresponds to mean-field rotations, {\it i.e.} without qubit entanglement.
Qubits were selected based on ensuring the one gate fidelity of the qubits 
were greater than 0.99. 

After the BK transformation, terms of the resulting Hamiltonian that form a mutually commutative 
set of operators are grouped together to perform a measurement on all of them at the same time. 
Since all the operators within a commutative set share eigenfunctions this procedure reduces the 
spread of measurement results due to general non-commutativity of the BK Hamiltonian and its terms. 

The expectation value of each commutative group was obtained by averaging over 1000 and 10000 
measurements for H$_2$ and H$_2$O, respectively.  
A post-processing procedure removing results with incorrect electron numbers and spin evaluated for each read 
was used for these measurements. It is shown below that this post-processing only removes results that 
appear due to experimental noise and does not alter quantum distributions of measurements.   
Upon assembling the expectation value of the total Hamiltonian from the expectation values of commutative 
groups, this procedure was repeated 20 (4) times for H$_2$ (H$_2$O) 
to obtain representative statistics for the Hamiltonian 
expectation values. Averages and standard deviations calculated over these 20 (4) 
Hamiltonian expectation values are reported as the final averages and standard deviations obtained on the QPU. 
Since our time was limited on the Rigetti system, we were not able to perform a more in-depth sampling procedure. 
All four experiments were performed in 24 hours over the course of 5 sessions. 

The classical optimization step for the VQE, we implemented the Nelder-Mead (NM) algorithm.\cite{NM:1965} 
Methods such as conjugate gradient descent were tried, but the NM algorithm demonstrated 
more robustness to the noise that was generated by errors. 

\subsection{Post-processing procedure}

\paragraph{Description of the procedure.} For the illustration purpose, let us assume that our 
Hamiltonian has two non-commuting 
parts, $\hat H = \hat A + \hat B$, $[\hat A,\hat B]\ne 0$ ({\it e.g.} $\hat A$
and $\hat B$ are two non-commuting Pauli words). Estimating the average of $\hat H$ 
on a wavefunction $\ket{\Psi}$ is done by adding the averages from non-commuting parts, 
$\braket{\Psi\vert\hat H\vert\Psi}= \braket{\Psi\vert\hat A\vert\Psi}+\braket{\Psi\vert\hat B\vert\Psi}$. The 
averages for both parts are computed by doing repetitive measurements, but these measurements 
collapse $\ket{\Psi}$ to different eigenfunctions because 1) $\hat A$ and $\hat B$ do not share eigenfunctions, 
and 2) $\ket{\Psi}$ is not generally an eigenfunction of $\hat A$ or $\hat B$. If we denote the eigenfunctions and 
eigenvalues of $\hat A$ ($\hat B$) as $\ket{f_n}$($\ket{g_n}$) and $a_n(b_n)$, respectively, then the Hamiltonian 
average is 
\bea
\braket{\Psi\vert\hat H\vert\Psi}&=&\sum_n a_n|\braket{\Psi\vert f_n}|^2 + b_n|\braket{\Psi\vert g_n}|^2.
\eea 
The probabilities $|\braket{\Psi\vert f_n}|^2$ and $|\braket{\Psi\vert g_n}|^2$ are not measured but instead 
emerge as a result of collecting results of individual measurements of $\hat A$ and $\hat B$. 
The post-processing procedure simply removes the eigenvalues $a_n$ and $b_n$ if corresponding 
eigenfunctions $\ket{f_n}$ and $\ket{g_n}$, available as readouts of QPU, 
violate the correct number of electrons (or any other known good quantum number). 

\paragraph{Invariance of the quantum average to the procedure.} 
The post-processing procedure based on the electron number operator 
is equivalent to introducing a projector $\hat P_N=\ket{N}\bra{N}$ to the eigen-subspace 
corresponding to the correct number of electrons, $\hat N\ket{N} = N\ket{N}$,
\bea\notag
\braket{\Psi\vert\hat P_N\hat H\hat P_N \vert\Psi} &=&
\sum_n a_n|\braket{\Psi\vert\hat P_N\vert f_n}|^2 \\
&&+ b_n|\braket{\Psi\vert\hat P_N\vert g_n}|^2.
\eea  
Indeed, eigenvalues associated with $|\braket{\Psi\vert\hat P_N\vert f_n}|^2$ and 
$|\braket{\Psi\vert\hat P_N\vert g_n}|^2$ will not contribute if corresponding eigenfunctions are 
orthogonal to the $N$-subspace: $\hat P_N\ket{f_n}=0$ and $\hat P_N\ket{g_n}=0$. On the other hand,
assuming that $\ket{\Psi}$ is within the $N$-subspace, $\bra{\Psi}\hat P_N = \bra{\Psi}$, it is straightforward
to see that $\braket{\Psi\vert\hat P_N\hat H\hat P_N \vert\Psi} = \braket{\Psi\vert\hat H \vert\Psi}$, and hence,
the Hamiltonian expectation values are not affected by the post-processing procedure. 

\paragraph{Noise reduction.} To understand how the post-processing reduces the noise related to uncontrolled
qubit interactions with its environment, it is convenient to employ the density matrix formalism. Let us 
denote the ideal pure-state density as $\rho_0 = \ket{\Psi_0}\bra{\Psi_0}$, while the real mixed-state
density, appearing due to spurious interactions, is $\rho = \sum_{i=0} \omega_i \ket{\Psi_i}\bra{\Psi_i}$.
Some components associated with the noise ($\{\ket{\Psi_i}\}_{i\ne 0}$) 
violate the correct value for the number operator, and therefore, $\hat P_N \ket{\Psi_i} =0$,
we will refer to the associated part of the density as reducible, $\rho_{R}$, while the rest of $\rho$
will be referred as irreducible, $\rho_{I}=\hat P_N \rho \hat P_N$. Clearly, substituting the Hamiltonian 
average with $\rho$, $\bar{E} = {\rm Tr}[\rho \hat H]/{\rm Tr}[\rho]$, by that with 
$\rho_I$, $\bar{E}_I = {\rm Tr}[\rho_I \hat H]/{\rm Tr}[\rho_I]$ makes results more accurate because of the 
noise reduction. By repeating the arguments shown for the pure state consideration, 
it is straightforward to show that the post-processing procedure removes the reducible 
part of the density in the Hamiltonian average, and hence it produces a more accurate average.

\subsection{Acknowledgements}

The authors thank P. Brumer, M. Mosca, and V. Gheorghiu for stimulating discussions. 
The authors are grateful to the support that Will Zeng, Nick Rubin, and Ryan Karle provided 
in terms of discussion of applying the theory onto their hardware and for access to the 19Q-Acorn 
Quantum computer. OTI Lumionics Inc. would like to acknowledge the support of the Creative 
Destruction Lab provided in facilitating the interbusiness collaborations between OTI Lumionics Inc. and Rigetti Computing Inc. A.F.I. acknowledges financial support from Natural Sciences and 
Engineering Research Council of Canada through the Engage grant. 

%

\end{document}